# Anomalous spectral scaling of light emission rates in low dimensional metallic nanostructures

D. A. Genov[*1,2], R. F. Oulton[*1], G. Bartal[1], and X. Zhang[†1,3]

[1]NSF Nanoscale Science and Engineering Center, University of California,
3112 Etcheverry Hall, Berkeley, CA 94720
[2]College of Engineering and Science, Louisiana Tech University, Ruston, LA 71272
[3]Materials Sciences Division, Lawrence Berkeley National Laboratory,
1 Cyclotron Road, Berkeley, California 94720, USA
*These authors contributed equally to this work
[†]e-mail: xiang@berkeley.edu

**The strength of light emission near metallic nanostructures can scale anomalously with frequency and dimensionality. We find that light-matter interactions in plasmonic systems confined in two dimensions (e.g., near metal nanowires) strengthen with decreasing frequency owing to strong mode confinement away from the surface plasmon frequency. The anomalous scaling also applies to the modulation speed of plasmonic light sources, including lasers, with modulation bandwidths growing at lower carrier frequencies. This allows developing optical devices that exhibit simultaneously femto-second response times at the nano-meter scale, even at longer wavelengths into the mid IR, limited only by non-local effects and reversible light-matter coupling.**

## I. INTRODUCTION

The search for ultrafast, compact and efficient light sources drives contemporary photonics toward new regimes where light-matter interactions are strongly enhanced. While the orders of magnitude discrepancy between electronic length scales and the wavelength of the light imposes spatial and temporal limitations, the emission of optical devices can be enhanced by diminishing the photonic mode volume and increasing the photonic density of states (DOS) [1]. For example, dielectric based pillar [2,3], photonic crystal [4,5] and microdisc [6,7] cavities have emerged as the means to achieve substantial enhancements in spontaneous emission, stimulated Raman scattering and non-linear frequency conversion. Having reached near diffraction limited sizes, the main route to further enhancement is through the cavity quality factor (Q-factor) [2-7]. While this approach is promising for some applications, it increases the photon lifetime in the cavity, thereby limiting the response time and bandwidth of the emitter-cavity system [5].

Surface Plasmons (SPs), the quasi-particles arising from the coupling of light with electrons at metal-dielectric interfaces opens the possibility to enhance light-matter interactions even with low DOS (or low Q- factor) by accessing mode sizes below the diffraction limit that are closer to the length scales of electronic wavefunctions [8, 9]. Surface plasmons have been also used to observe naturally weak physical effects with enhanced sensitivity, such as fluorescence [10, 11] and Raman spectroscopy [12, 13], and very recently were investigated as potentially new types nanolaser systems and ultrafast light sources at optical and infrared frequencies [14-17].

This paper shows that the strength of light-matter interactions in low dimensional plasmonic systems manifests unique spatial and temporal scalings with frequency that are highly unusual when compared to conventional optics. We find that the light-matter interaction strength in plasmonic systems confined in two dimensions e.g., metal nanowires, scales anomalously with frequency and exhibits improved confinement and emission rates away from the surface plasmon frequency, despite the diminishing role of electron oscillations in the confinement. Strong light-matter interactions in such systems are shown to provide simultaneously femto-second and nano-meter regimes of operation, even at longer wavelengths into the mid IR. Furthermore, coherent SP sources [17, 18] are now viable with recent demonstration of SP lasers at optical and infrared frequencies [19-21].

The rest of this paper is organized as follows: in Sec. II we set the theoretical method used to calculate the SPs modal characteristics and interaction with dipolar emitters, in Sec. III we study the frequency scaling of mode confinements and spontaneous emission rates related to SPs with various degrees of confinement, in Sec. IV we investigate a prospective Surface Plasmon Laser (SPL) and extract it relevant characteristics such as modulation bandwidths. The effects of SP mode dimension, cavity Q-factors and carrier frequency are addressed. Conclusions are provided in Sec. V.

## II. MASTER EQUATION: SURFACE PLASMONS MODAL VOLUMES AND SPONTANIOUS EMISSION RATES

We analyze the electromagnetic interaction between a two level dipole emitter in close proximity to metal-dielectric systems exhibiting mode confinement in one, two and three dimensions, shown in Fig. 1. A planar metal-dielectric interface [22-24] and a thin metal nanowire [25-27] represent plasmonic systems with one (1D) and two (2D) dimensional confinement, respectively, where collective electron oscillations are bound to the surface or wire but are free to propagate in the remaining unbound directions as surface plasmon polaritons (SPPs). These systems are non-resonant and support SPPs at energies below the plasma frequency. On the other hand, a metal nanoparticle supports a resonant localized Surface Plasmon (SP) confined in all three dimensions (3D) [28].

For the sake of simplicity we assume that the 3D SP resonance frequency $\omega_{sp}$ is tuned to that of the emitter, $\omega$. In general, the emitter couples to an electric field composed of a variety of modes from unconfined (0D) radiation to cavity modes confined in all 3 dimensions. The emitter-field interaction, under the rotating wave approximation [29], is then described by the master equation:

$$\dot{\rho} = \frac{1}{i\hbar}\sum_D [\widehat{\mathcal{H}}_D, \rho] + \sum_D \gamma_D \hat{\mathcal{L}}(\hat{a}_D)\rho + \gamma_{nr}\hat{\mathcal{L}}(\hat{\sigma}_\pm)\rho, \qquad (1)$$

where $\widehat{H}_D = i\hbar g_D\left(\hat{\sigma}_-\hat{a}_D{}^\dagger - \hat{\sigma}_+\hat{a}_D\right)$ is the interaction Hamiltonian describing the coupling between emitter and SP modes confined in D dimensions; $\gamma_{nr}$ describes non-radiative emitter loss, $\gamma_D$ describes loss for mode confined in D dimensions; $\rho$ is the density matrix; $\hat{a}^\dagger(\hat{a})$ and $\hat{\sigma}_\pm$ are the SP creation (annihilation) and Pauli spin-flip operators, respectively; and $\hat{\mathcal{L}}(\hat{A})\rho = (2\hat{A}\rho\hat{A}^\dagger - \hat{A}^\dagger\hat{A}\rho - \rho\hat{A}^\dagger\hat{A})$ is the Liouville operator describing the system's irreversible losses. In the following, we examine the partial emitter-field interactions for each dimensionality separately. The emitter-mode coupling strength $g_D$ depends on the overlap of electronic wave-functions and the local electric field, $\boldsymbol{E}(\boldsymbol{r}_a)$, and for emission averaged over all dipole orientations is given as

$$g_D(\boldsymbol{r}_a) = \frac{1}{\hbar}\langle 1|\boldsymbol{d}.\boldsymbol{E}(\boldsymbol{r}_a)|2\rangle = \frac{1}{\pi}\left(\frac{2\omega\Gamma C_D f(\boldsymbol{r}_a)}{\xi_{3-D}}\right)^{1/2}\left(\frac{\pi c}{n_a\omega}\right)^{(3-D)/2}, \qquad (2)$$

where $\boldsymbol{d}$ is the dipole moment, $\boldsymbol{r}_a$ is the dipole position and $\Gamma = n_a\omega^3|\boldsymbol{d}|^2/3\pi\varepsilon_0\hbar c^3$ is the natural spontaneous emission rate of the emitter in an open lossless medium of refractive index $n_a$. To avoid rapid non-radiative decay of emitters in close proximity to metal surfaces we set the position dependence of the coupling strength, $f(\boldsymbol{r}_a) = u_E(\omega, \boldsymbol{r}_a)/\text{Max}\{u_E(\omega, \boldsymbol{r})\} = e^{-1}$, where $u_E(\omega, \boldsymbol{r})$ is the modal electrical energy density [18, 26]. In equation (2), the $\xi_{3-D}$ is a quantization (3-D)-volume accounting for the unbound dimensions and the confinement factor is,

$$C_D = \frac{\text{Max}\{2u_E(\boldsymbol{r})\}}{\int d^D \boldsymbol{r} u_{EM}(\boldsymbol{r})} \left(\frac{\pi c}{n_a \omega}\right)^D. \quad (3)$$

Here, the electromagnetic energy density, $u_{EM}(\omega, \boldsymbol{r}) = u_E(\omega, \boldsymbol{r}) + u_M(\omega, \boldsymbol{r}) = (d\omega\varepsilon(\omega)/d\omega)|\boldsymbol{E}(\omega, \boldsymbol{r})|^2/2 + \mu_0|\boldsymbol{H}(\omega, \boldsymbol{r})|^2/2$, is normalized to the vacuum energy $\xi_{3-D} \int d^D \boldsymbol{r} u_{EM}(\omega, \boldsymbol{r}) = \hbar\omega$.

In this paper we will utilize the solutions of the master equation modified only through the loss rates and coupling coefficients as expressed in Eq. (1) [30]. The atom-cavity dynamic involves both reversible and irreversibly emission by the atom, where energy is exchanged between light and matter states at the Rabi flopping frequency $\Omega_0 = \sqrt{-\mu_-^2 + |g_0|^2}$ with a transition rate $\Gamma_T = -\mu_+ \pm i\Omega_0$, where $\mu_\pm = (1/2)(\Gamma_I \pm \gamma_c)$; $\Gamma_I = \sum_{D\neq 0} \Gamma_D + \gamma_{nr}$ is the decay rate due to all the irreversible loss channels; and $\gamma_c = \gamma_{D=0}$ is the cavity loss rate. The high losses, encountered in most SP systems, $\gamma_c \gg \Gamma_I$, result in an irreversible coupling regime, with an emission rate into the SP mode of $|g_0|^2/\gamma_c$ without vacuum Rabi Oscillation.

In the absence of cavity feedback, the emitter couples irreversibly to a continuum of modes, which allows the use of Fermi's Golden rule to describe the emission rate, $\Gamma_D = \pi/2|g_D|^2 \xi_{3-D} G_D$, where $G_D$ is the density of states (DOS) of modes confined in $D$ dimensions. For free space radiation modes ($D = 0$), the DOS is $G_0 = n_a^3 \omega^2/\pi^2 c^3$, hence $\Gamma_0 = \Gamma$. Increasing the confinement dimensionality modifies the DOS: $G_1 = n_p n_g \omega/2\pi c^2$ for confinement in 1D (propagation on a metal plane) and $G_2 = n_g/\pi c$ for confinement in 2D (propagation along a metal wire), where $n_p$ and $n_g$ are respectively the phase and group indexes of SP modes (see Auxiliary information). The partial spontaneous emission rate enhancements are therefore $F_D = \Gamma_D/\Gamma = C_D G'_D$, where we have introduced the normalization,

$$G'_D = \frac{\omega G_D}{\pi}\left(\frac{\pi c}{n_a \omega}\right)^{3-D}. \quad (4)$$

We note that this is similar to normalizing the SP DOS to the DOS of non-dispersive modes of effective index $n_a$ with the same confinement dimensionality. However, this is only true for $D = 0$, where $G'_0 = G_0(n_p(\omega))/G_0(n_a)$. For $D = 1$, $G'_1 = G_1(n_p(\omega))/2G_1(n_a)$ and for $D = 2$, we find $G'_2 = G_2(n_p(\omega))/\pi G_2(n_a)$. The additional factors account for the reduction in available modes for low dimensional modes compared to free space. As evident from DOS (equation 4), plasmonic systems confined in 1 and 2 dimensions are capable of emission enhancements over a broad range of frequencies without the need of a cavity [22,23].

Conversely, 3D confined SP systems act as effective cavities and exhibit a resonance condition and must be tuned to match the emitter's spectrum. Furthermore, the optical feedback onto the emitter can introduce both irreversible and reversible energy exchange between light and matter states. However, as pointed above the high metal loss, encountered in most SP systems, $\gamma_3 \gg \gamma_{nr} + \sum_{D \neq 3} \Gamma_D$, results in irreversible coupling with a total emission rate $\Gamma_T = \gamma_{nr} + \sum_{D \neq 3} \Gamma_D + |g_3|^2/\gamma_3$. The emission rate enhancement for the cavity mode is therefore $F_3 = C_3 G'_3$, where the 3D cavity DOS, $G_3 = 2/\pi\gamma_3 = 2Q_{sp}/\pi\omega$, is the peak value of the cavity lineshape function. Substituting the modal confinement $C_3 = (\pi c/n_a \omega)^3 V^{-1}$, we recover the well-known Purcell enhancement factor for a cavity with volume $V$ and quality factor $Q_{sp}$, $F = 2\pi(c/n_a\omega)^3 Q_{sp}/V$ [1].

## III. FREQUENCY SCALING OF PLSMONIC SYSTEMS WITH VARIOUS DEGREES OF CONFINEMENT

A unique spectral scaling of dipole emission rates into surface plasmons modes arises from the underlying dependence of their optical density of states (DOS) and mode confinement factors on frequency, as shown in Fig. 2. In these calculations, we consider air/silver configurations for different system dimensionalities. To describe the metal permittivity we use the Drude model, $\varepsilon_m(\omega) = \varepsilon_b - \omega_p^2/(\omega^2 - i\omega\omega_\tau)$, where $\varepsilon_b = 5$ is the contribution due to bound electrons, $\hbar\omega_p = 9.1$ eV is the bulk plasma frequency, and $\hbar\omega_\tau = 21$ meV is the relaxation rate due to electron-phonon scattering [31]. In 1D and 2D confined plasmonic systems, the intrinsic densities of states dramatically increase near the surface plasmon frequency, $\omega_{sp}$, where a larger proportion of the modal energy resides in the dispersive metal. The redistribution of the modal energy from the metal into the dielectric at lower frequencies results in a DOS similar to that of free photons. While the DOS of both 1D and 2D confined surface plasmon polaritons (SPPs) scale similarly with frequency, the spectral dependences of their confinement factors exhibit opposing trends. 1D SPPs weakly confined to metal-dielectric interfaces for $\omega \ll \omega_{sp}$, have a confinement factor that scales as $C_1 = 2\pi n\omega/\omega_p < 1$, indicating that such SPPs extend to sizes substantially larger than the diffraction limit. However, the most intriguing behavior is found for SPPs confined in 2D, whose mode confinement factor scales anomalously with frequency and actually rapidly increases at low frequencies. In contrast to exponential decay of plasmonic modes confined in 1D, the electric field of 2D confined plasmonic modes decays $\propto r^{-1}$ near the nanowire and $\propto \exp(-\kappa r)r^{-1/2}$ for large distances from the nanowire, $r \gg \lambda$. Consequently, the mode confinement area is related to the nanowire's radius, $a$, when $a/\lambda \ll 1$, and is independent of the wavelength, due to the $r^{-1}$ near-field decay [26]. This peculiar phenomenon can be physically explained in terms of electromagnetic energy storage in surface electron oscillations screened by the core of the metal nanowire. When the nanowire is much narrower than the wavelength, the electromagnetic field associated with these charge oscillations is geometrically constrained to scale inversely with distance. A significant portion of the SP mode energy therefore resides near the metal-dielectric interface over a broad range of frequencies, similar to the charge storage in a capacitor. The confinement factor, $C_2 \approx \pi(c/\omega a)^2$, therefore intensifies strongly with diminishing frequency and nanowire radius. In is important to mention that apart of the fundamental nanowire mode ($m = 0$) studied here, one may also consider higher order modes. However, as shown by Chang *et al.* [27], these modes experience either a cutoff for a given nanowire radius $a$ or for $m = 1$ have modal volume that exponentially increases with decreasing $a$. Thus, the fundamental mode is the mode of choice should one pursue fast optical response and small modal volumes. We note that anomalous scaling with frequency may also be found in other types of metallic nanostructures, e.g., two parallel metal planes separated by a dielectric gap exhibit a more moderate scaling of the confinement factor as $\omega^{-1}$.

The limits of nano-plasmonics are manifested in 3D-confined systems, i.e., metal nano-particles [17, 28]. The smallest nanoparticles exhibit very little radiative loss [32] and are well understood by a quasi static description where the particle's shape determines its resonance frequency, while its density of states is set by the metal's permittivity alone; $G_3 = 2/\pi\gamma_3 = \text{Re}[d\varepsilon_m/d\omega]/\pi\text{Im}[\varepsilon_m]$ [33]. However, inhibitive metal loss due to non-local phenomena, such as Landau damping and the anomalous skin effect, can further limit both DOS and mode confinement for very small particle sizes [34].

We define the limit of plasmonics at the breakdown of the continuous theory of metals, when the permittivity becomes dependent on the spatial frequency $k_a$. This naturally introduces a minimal particle size, $a = 2\pi/k_a < v_F/\omega$, where $v_F$ is the Fermi velocity, through the constraint $\delta = |\varepsilon_m(\omega, k_a)/\varepsilon_m(\omega) - 1| \ll 1$. We describe the non-local permittivity using the Klimontovich-Silin-Lindhart formula [34]. To estimate the Purcell factor for the resonant SP modes (3D confinement) we again use the general result for the emission rate enhancement $F_3 = C_3 G_3' = 2\pi^{-2} C_3 Q_3$, where $Q_3 = \omega/\gamma_3$ is the Q-factor of the SP resonances (we tune the particle morphology to be at resonance for the operation frequency $\omega$ ) [33]. Finally, in the calculations we set the particle size such that $k_a v_F/\omega = 0.13$ and the deviation in Eq. (5) from the bulk metal permittivity is less then 1%.

The limit of non-local effects manifested in 3D-confined (resonant) systems is illustrated in Figs. 2a and 2b with solid lines, where we span the frequency range by modifying a nanoparticle's shape to tune its resonance frequency. At low frequencies, the mode confinement is given as $C_3 \approx 12\pi^2 (c/a\omega)^3$, and at the onset of nonlocal effects the particle size must scale as $a \approx v_F/\omega$ where $v_F$ is the Fermi velocity. Concurrently, the mode confinement saturate at a finite value proportional to $(c/v_F)^3$. At high frequencies, we observe a different behavior with the mode confinement decreasing drastically due to the metal's transparency $\varepsilon_m \to 1$.

The intrinsic characteristics of SPs alone can lead to dramatic enhancements in the spontaneous emission rates (Purcell effect) of emitters near plasmonic systems as shown by the solid lines in Fig. 3. In general, the enhancement factor is proportional to the product of the DOS and mode confinement factors discussed above and shown in Figs. 2(a) and 2(b), respectively. As the DOS of 2D confined systems maintains a nearly constant value at low frequencies, it is the anomalous scaling of the mode confinement with frequency that drives the scaling of the spontaneous emission rates. Namely, the rate of spontaneous emission in 2D confined systems rapidly increases at lower frequencies, in sharp contrast to plasmonic systems confined in one dimension and the onset of non-local effects in 3D confined SPs.

Light matter interactions of 1D and 2D confined SPPs can be further enhanced by introducing a cavity, where the interference between multiple reflections within the cavity can substantially modify the intrinsic DOS. The emission rates will therefore be further enhanced by a factor proportional to the cavity *finesse*, $\mathcal{F} = Q/n = \lambda Q/2 n_g L$, where $n$ is the mode order, in exchange for a modest reduction of the bandwidth. Namely, an emitter near 1D SPPs resonating between appropriately spaced mirrors will emit at a rate $|g_{c,2}|^2/\gamma_c = 2\Gamma_2 \mathcal{F}/\pi$ and for 2D SPPs in a square cavity of side $L$ the emission rate is $|g_{c,1}|^2/\gamma_c = 4\Gamma_1 \mathcal{F}/\pi^2 \sqrt{n^2 + 1}$, where $g_{c,D}$ and $\gamma_c$ are the generalized coupling coefficient and loss rate for cavity enhanced SPPs confined in D dimensions. In both 1D and 2D confined systems, additional confinement is also possible near the surface plasmon frequency, where the increased phase index, $n_p$, of the SPPs allows a reduction in the cavity length. However, when cavities are longer than the SP propagation distance, $\mathcal{F} \to 0$, we recover the non-resonant intrinsic emission rates.

The broken lines in Fig. 3 show the emission rate enhancement of 1D and 2D systems with first order cavities added, i.e. $m = 1$. Here, reduced mode sizes and moderate Q-factors, limited by the intrinsic dissipation ($R_c = 99\% \gg e^{-\alpha L}$, where $\alpha = 2(\omega/c)\text{Im}(n_p)$ is the SP's dissipation along the cavity length $L$), allow for improved emission characteristics. Spontaneous emission rates are ultimately limited by the onset of reversible light-matter coupling and cannot exceed half the intrinsic cavity loss rate, when $|g_{c,D}| > \gamma_c = (c/2n_g L)\ln(1/R_c e^{-\alpha L})$ [29,30]. While the Purcell factors in 3D confined SP systems approach six orders of magnitude at the onset of non-local effects, reversible light matter coupling is likely to be the limiting factor, where $|g_3| > \gamma_3$. Nevertheless, this is sufficient to access sub-picosecond spontaneous emission lifetimes.

**IV. SURFACE PLASMON LASER (SPL)**

Emission in strongly confined plasmonic systems is typically dominated by intrinsic SPP loss [28,32,35] inhibiting them from realizing practical light sources. However, the feedback of 1D/2D confined SPPs in cavities and 3D confined SPs can induce stimulated emission resulting in nano-plasmonic lasers with restored out-coupling efficiency and coherence [19-21]. In what follows, we examine how the spectral scaling of both mode confinement and density of states impact the temporal dynamics of plasmonic laser systems. While it is widely accepted that the speed of plasmonic lasers based on 3D confinement is governed by the short passive lifetimes of SPs [14], here we identify a new regime where the speed of 2D-confined plasmonic lasers is governed by the anomalous scaling of mode confinement with frequency. This finding makes SPPs confined in just two dimensions extremely favorable for fast laser systems without relying on high gain to compensate high mode loss.

Laser action of SPPs is described by the master equations for an ensemble of four level emitters with rapidly depleting ground state coupled to a single cavity mode under the assumption of fast emitter-photon de-phasing dominated by SP cavity loss, $\gamma_c$, such that,

$$\dot{N} = J - \Gamma_T N(1 + \beta S),$$
$$\dot{S} = \beta \Gamma_T N(1 + S) - \gamma_c S, \tag{5}$$

where $\beta = |g_{c,D}|^2/\gamma_c \Gamma_T$ is the spontaneous emission factor - the probability that emission couples into the cavity mode; $J$ and $N$ are the pump rate and population inversion, respectively; and $S$ is the number of SPP quanta inside the cavity. These expressions are valid within the weak coupling limit, ($\gamma_c \gg \beta \Gamma_T$).

Nano-scale SPP lasers differ from their conventional counterparts in their ability for threshold reduction. The threshold pump rate is the ratio of the cavity loss rate and the spontaneous emission factor, $J_{th} = \gamma_c/\beta$. While conventional lasers, having $\beta \ll 1$, require low cavity losses to ensure realistic thresholds, SPP lasers can operate at much higher cavity losses owing to their high $\beta$-factor ($\beta \approx 1$). The high $\beta$ stems from the increased light matter coupling strength, inducing enhancements of both spontaneous and stimulated emission rates beyond that available to diffraction limited light. This opens the possibility to achieve laser action of SPPs with realistic pump rates despite high cavity loss and the capability for very fast response times.

To explore the temporal characteristics of SP lasers, we examine their response to small signal modulations, $S = S_0 + s(\omega)e^{i\omega t}$, $J = J_0 + j(\omega)e^{i\omega t}$ and $N = N_0 + n(\omega)e^{i\omega t}$, where $S_0$, $J_0$, and $N_0$ are the steady state SPP mode number, pump rate, and population inversion, respectively. The SPL response to the pump modulation in first order of perturbation is described by the spectral response function

$$\Theta(\omega) = \left|\frac{s(\omega)}{j(\omega)}\right| = \frac{\beta \Gamma_T (1+S_0)}{\sqrt{(\omega^2 - \omega_r^2)^2 + \omega^2 \omega_p^2}} \qquad (6)$$

where $\omega_p = \gamma_c + \Gamma_T(1 - \beta N_0 + \beta S_0)$ and $\omega_r^2 = \Gamma_T(\gamma_c(1+\beta S_0) - \beta(1-\beta)\Gamma_T N_0)$ are related to the resonance frequency $\omega_{res} = \sqrt{\omega_r^2 - \omega_p^2/2}$. Fig. 4(a) depicts the response functions for 1D and 2D confined SPPs of Semiconductor/Silver nanostructures at the telecoms wavelength $\lambda = 1.5 \mu m$. The time response of the laser is characterized by the modulation bandwidth, $f_{3dB}$, defined as the frequency at which the response function decays to half of its zero-frequency value. For SP cavities, one has $\beta \approx 1$ and at low pump rates, $r = J_0/J_{th} > 1$, the modulation bandwidth is limited by relaxation oscillations, $f_{3dB} \leq \sqrt{3}\omega_r/2\pi$, where $\omega_r = \sqrt{\Gamma_T \gamma_c r} < \gamma_c$. At high pump rates, $r \gg 1$, a damping dominated response occurs ($\omega_r > \gamma_c$), and we have $f_{3dB} \leq \sqrt{3}\gamma_c/2\pi$. In general, we find that for high loss cavities, inherent to SP systems, the 3dB bandwidth can be completely described through a single universal parameter

$$f_{3dB} = \frac{\gamma_c}{\pi}\Omega(\omega_r/2\gamma_c), \tag{7}$$

where $\Omega(\zeta) = \zeta\left(1 - 2\zeta^2 + 2\sqrt{1 + \zeta^2(\zeta^2 - 1)}\right)^{1/2}$, and the laser's maximum response rate depends on pump rate and is in the range, $\omega_r < 2\pi f_{3dB}/\sqrt{3} < \gamma_c$. The modulation capability of the SPL over visible and near IR frequencies, as shown in Fig. 4 (b), exemplifies why plasmonic systems confined in two dimensions are highly suitable for such devices. Figure 4(b) shows the improvement in the bandwidth of SPP lasers compared with conventional ones with the same cavity feedback, $f_{3dB}^{sp}/f_{3dB}^{cav}$. While at high frequency the short passive lifetimes of plasmonic cavity modes govern the modulation speed, at low frequency and for nominal pump rates, $r$, it is the confinement that is important, $2\pi f_{3dB}^{sp}/\sqrt{3} = \sqrt{\Gamma_T\gamma_c r} \approx \sqrt{2r\omega C_3\Gamma}/\pi$, and the modulation speed is independent of the passive cavity mode lifetime. For high pump rates, $r$, the modulation speed continues to increase up to the cavity loss rate $2\pi f_{3dB}^{sp}/\sqrt{3} < \gamma_c$. For nominal pump power, we find that the maximum modulation speed inherits the anomalous scaling of SPPs confined in 2D and scales as $\sim\omega^{-1/2}$. This behavior suggests that fast plasmon lasers are possible without the need to rely on short passive SPP lifetimes and correspondingly high gain to achieve lasing. By minimizing the cavity losses in this way, more gain will be available for driving such lasers faster. On the other hand 1D confined SP lasers have a reduced bandwidth due to diminishing mode confinement at longer wavelengths, hence these systems are only advantageous near the surface plasma frequency, where passive mode loss is high, and the corresponding large gain requirement for lasing makes high bandwidth 1D confined SP lasers difficult to achieve [16].

In practical SP lasers the cavity losses should be as low as possible with an enhancement factor that scales as, $f_{3dB}^{sp}/f_{3dB}^{cav} = \pi^{-1}\sqrt{2C_3Q}$. While SP lasers with high cavity loss rates could be modulated faster, these devices would ultimately be limited by high thresholds and the gain available in natural materials. For example, under realistic conditions where gain suppression effects are included, the maximum modulation bandwidth cannot exceed, $f_{3dB} < \sqrt{3}\Gamma_T/2\pi + (3/4\pi^2)^{3/2}(\lambda^3/\epsilon n_a^3)Q\Gamma$, for $\gamma_c > \Gamma_T$, where $\epsilon$ is a gain suppression factor [36]. For typical gain materials at telecoms wavelengths, $\epsilon = 10^{-23}$ m$^3$, $\Gamma = 1$ GHz and $n_a = 3.5$ which, along with $Q = 20$, gives an upper bandwidth limit of $f_{3dB} < \sqrt{3}\Gamma_D/2\pi + 3.3$ THz. Gain suppression is likely to limit the response time of SPP lasers to < 10 THz.

## V. CONCLUSION

In conclusion, we have shown that SPs confined in 2D exhibit anomalous mode confinement that rapidly increases with decreasing frequency away from the surface plasmon frequency. This enables ultra small mode sizes even at telecomm and mid-IR wavelengths which, along with their coherent and guided radiation, makes 2D confined plasmonic systems ideal candidates for future laser devices. The anomalous spectral scaling of mode confinement also extends to the speed of plasmonic lasers where modulation bandwidths can exceed those of spontaneous emission based light sources, limited only by reversible light-matter coupling. Plasmonic lasers can bridge the size-gap between electronics and optics to deliver optical energy to nanometer length scales on femtosecond time scales, hence providing improved information throughputs and substantial size reduction of optical light sources. This unprecedented control over the speed of light-matter interactions holds the potential to greatly impact the development of highly compact optoelectronic circuits and ultrafast signal processing devices as well as introduce a new regime of intense field laser physics.

**ACKNOWLEDGMENTS**

This work has been supported by DOE under Grant No. 32-4001-56176, and Louisiana Board of Regents under contract number LEQSF (2007-12)-ENH-PKSFI-PRS-01.

**FIGURES**

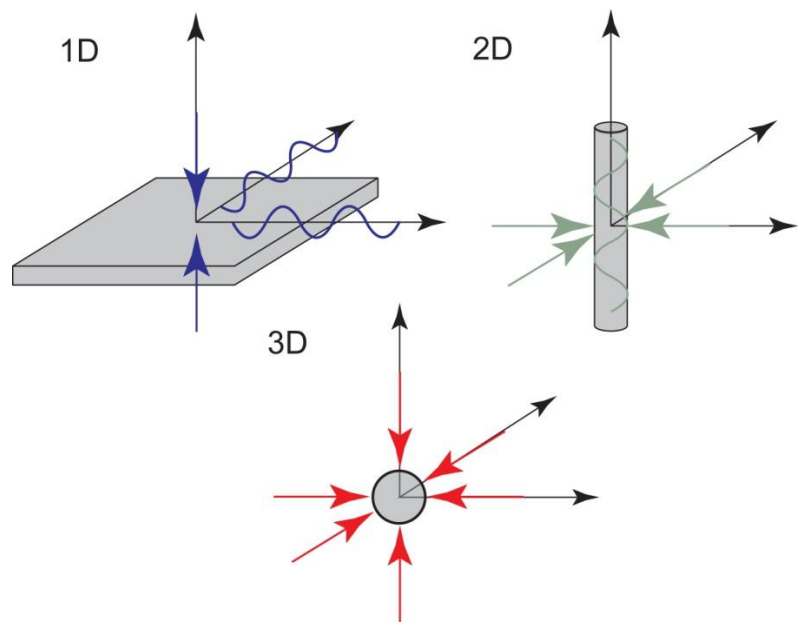

FIG. 1. (color online) Metal nanosctructures supporting surface plasmon modes with varying degrees of confinement. The 1D system is planar Silver/Air interface supporting surface plasmon polaritons (SPP) confined in one dimension. The 2D system is a 40 nm diameter Silver nanowire in air and supports the symmetric zero-order SPP confinement in two dimensions. The 3D system (red) corresponds to a nanosize Silver particle confined in all three dimensions.

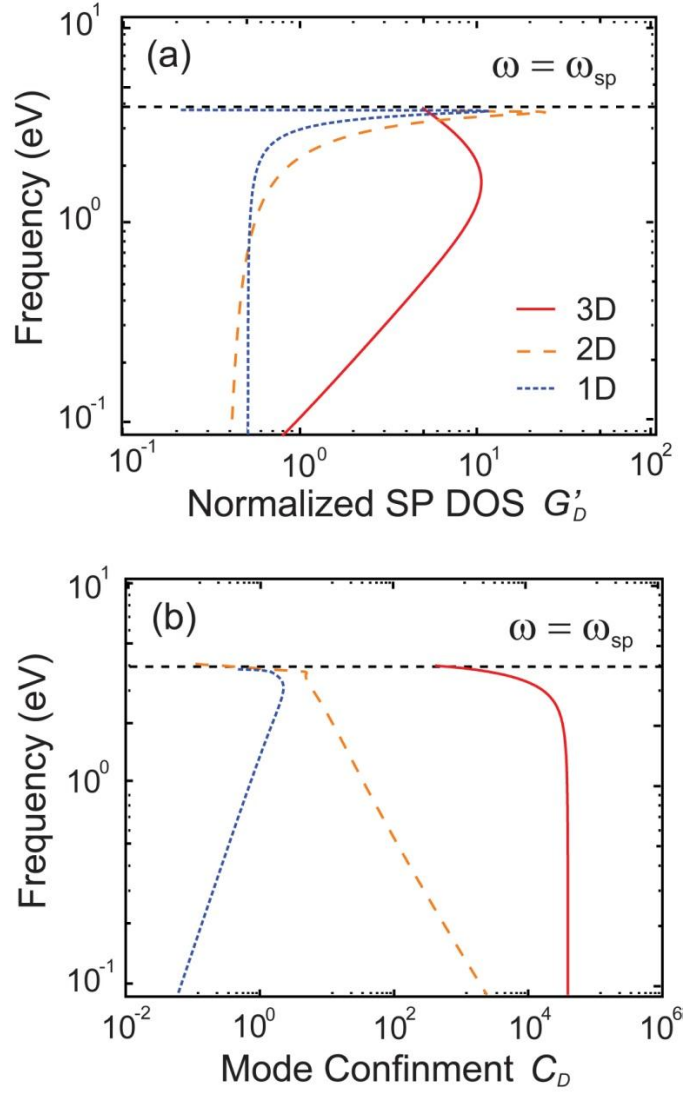

FIG. 2. (color online) Intrinsic characteristics of the Surface Plasmons (SPs). (a) Normalized densities of states, $G'_D$, and (b) effective mode confinement factors, $C_D$. While the 1D (dotted) and 2D (dashed) lines show the response for a single calculated structure, the red line (solid) identifies the onset of non-local effects in metal nanoparticles ($\delta = 0.01$) which have been morphologically tuned to provide localized Surface Plasmon resonances at the sampled frequencies.

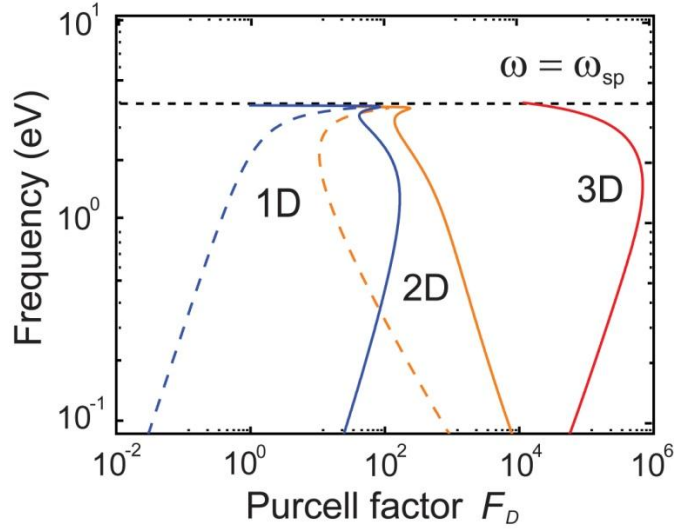

FIG. 3. (color online) Spontaneous emission rate enhancement for emitters coupled to SPs with confinement in a varying number of dimensions. This is the product of the confinement factor, $C_D$, and normalized DOS, $G'_D$, as shown in Figs. 2a and 2b. The responses for the 1D and 2D confined SPPs are divided into *intrinsic* (broken lines) and cavity-enhanced (solid lines) enhancement factors, where feedback has been introduced through cavity mirrors in a Fabry-Perot configuration with reflectivity, $R_c = 99\%$. The red line shows the limits of plasmonics where non-local effects become significant. This limit could be reached with metallic nano-particle SP resonators. In these calculations silver/air configurations are considered.

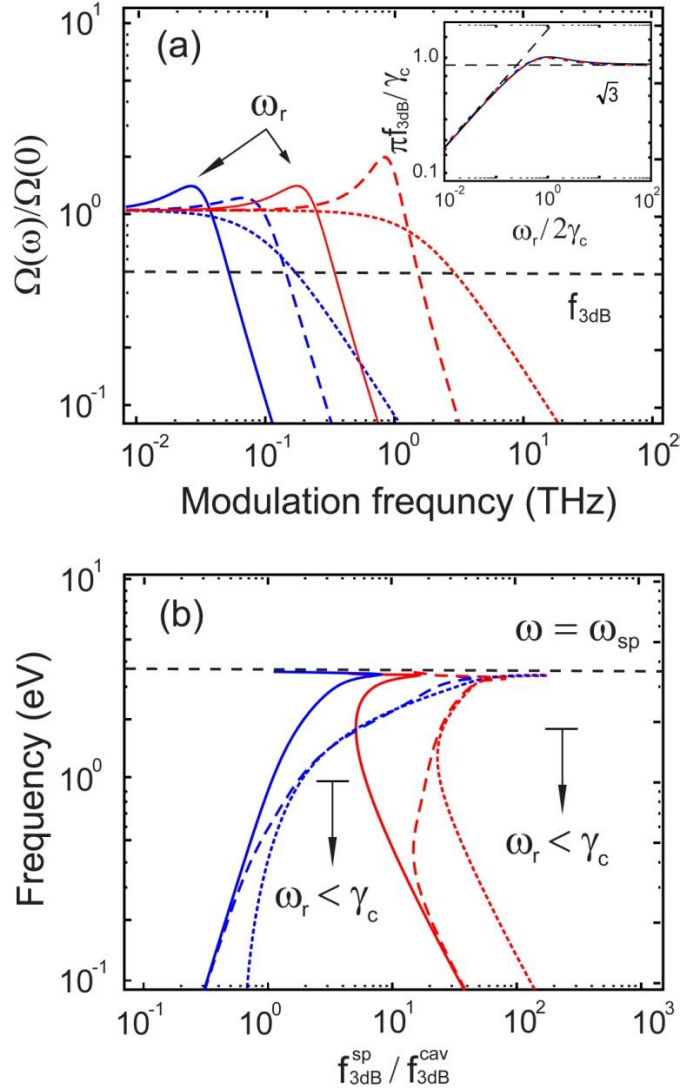

FIG. 4. (color online) Temporal response of Surface Plasmon Lasers. (a) The response functions for 1D (light curves) and 2D (dark curves) confined SPP cavities ($n = 1$, $R_c = 99.9\%$) for ideal four level emitters at telecoms frequency ($\omega = 0.83$ eV). We consider three different pump rates: $r = 10$ (solid lines), $r = 200$ (dashed lines) and $r \to \infty$ (dotted line). The 3dB modulation bandwidth follows a universal function (insert). At low pump rates, the bandwidth is limited by relaxation oscillations, $f_{3dB} = \sqrt{3}\omega_r/2\pi$, saturating at $f_{3dB} = \sqrt{3}\gamma_c/2\pi$, for high pump rates due to cavity damping. (b) The enhancement of the 3dB bandwidth compared to diffraction limited lasers in the optical and near infrared, for the same pump rates as in Fig. 4(a). A relaxation oscillation dominated response (solid lines) is observed at low pump rates and near the surface plasmon frequency where loss rates are high. A transition into cavity relaxation dominated response occurs when $\omega_r \gg \gamma_c$. In these calculations, the cavities' host permittivity is set to $\varepsilon = 2.25$.